\pdfoutput=1

\documentclass[fleqn, 12pt]{article}
\usepackage[utf8]{inputenc}
\usepackage[T1]{fontenc}
\usepackage{csquotes}
\usepackage{setspace}
\usepackage{authblk}
\usepackage{graphicx}
\usepackage{amsmath}
\usepackage{hyperref}
\usepackage[margin=1in]{geometry}
\usepackage{titlesec}
\usepackage[authoryear]{natbib}
\titleformat{\section}{\bfseries\uppercase}{\thesection.}{1em}{}
\titleformat{\subsection}{\bfseries}{\thesubsection.}{1em}{}
\titleformat{\subsubsection}{\itshape}{\thesubsubsection.}{1em}{}

\title{Small increases in agent-based model complexity can result in large increases in required calibration data}

\author[1,*]{Vivek Srikrishnan}
\author[2,1]{Klaus Keller}
\affil[1]{Earth and Environmental Systems Institute, Pennsylvania State University, University Park, PA, USA, 16802}
\affil[2]{Department of Geosciences, Pennsylvania State University, University Park, PA, USA, 16802}

\affil[*]{vxs914@psu.edu}
\date{}
\providecommand{\keywords}[1]{\textit{Keywords:} #1}

\begin{document}

\doublespacing

\flushbottom
\maketitle

\newpage

\begin{abstract}
Agent-based models (ABMs) are widely used to model coupled natural-human systems.  Descriptive models require careful calibration with observed data. However, ABMs are often not calibrated in a statistical sense. Here we examine the impact of data record structure on the calibration of an ABM for housing abandonment in the presence of flood risk. Using a perfect model experiment, we examine the impact of data record structures on (i) model calibration and (ii) the ability to distinguish a model with inter-agent interactions from one without. We show how limited data sets may not constrain a model with just four parameters. This indicates that many ABMs may require informative prior distributions to be descriptive. We also illustrate how spatially-aggregated data can be insufficient to identify the correct model structure. This emphasizes the need for utilizing independent lines of evidence to select sound and informative priors.
\end{abstract}

\keywords{agent-based modeling, statistical calibration, model selection}
\thispagestyle{empty}
\newpage

\pagenumbering{arabic}

\section{Introduction}

Agent-based models (ABMs) can be a useful tool for modeling and understanding how macro-scale/aggregate features of complex systems emerge from micro-scale/individual decisions, interactions, and feedbacks (``generative'' social science \citep{Epstein1999-fa}). As a result, they have found use in many application areas, including land use change \citep{Parker2003-fl, Evans2004-ra, Brown2005-fq, Evans2008-ym, Kelley2011-os, Evans2013-ov, Brown2017-tn}, ecology \citep{Black2012-bz, Grimm1999-ej, Van_der_Vaart2016-er}, flood risk \citep{Aerts2018-qz, Dubbelboer2017-dn, Haer2016-zm, Jenkins2017-tu, Tonn2017-ro}, and climate change adaptation \citep{Balbi2013-fe, Barthel2008-cu, Gerst2013-pe, Schneider2000-wa, Ziervogel2005-if}. 

Models can be designed to address different questions about the modeled system, including prediction, explanation, and demonstration \citep{Epstein2008-ko}. \citet{Marks2011-xu} propose a classification of simulation models as demonstrative or descriptive based on the model’s purpose. Demonstrative ABMs are used to illustrate that patterns of interest can be produced through local-level rules and interactions. Descriptive ABMs are intended to reproduce observed phenomena for the purpose of explanation, prediction, or both. The descriptive model category includes both simpler, ``strategic'' models, intended  and more complex, ``tactical'' models \citep{Holling1966-rl}. Early ABMs, such as the pioneering work on segregation \citep{Schelling1971-lv}, were primarily demonstrative \citep{Janssen2006-yy, Marks2011-xu}. Over time, there has been an increase in descriptive models \citep{Janssen2006-yy}, the most famous of which is arguably the Artificial Anasazi Model \citep{Dean2000-hr}.

Both demonstrative and descriptive models require tests to ensure that the model works as intended. Descriptive models also benefit from a comparison of model output against observations \citep{Marks2011-xu}. Simulations of hindcasts demonstrate that the model reproduces historical data, though this is not the same thing as demonstrating that the model is a reproduction of system dynamics \citep{Oreskes1994-tz}, as all models are approximations of real processes \citep{Box1976-vy}.

This observation that models are only capable of approximating, rather than reproducing real system dynamics, shows the importance of model calibration, the process of selecting model structures and parameter values, for descriptive modeling.  One common approach to calibration is to tune model parameters until model outputs are close to the empirical data \citep{Kelley2011-os, Schwarz2009-ry, Van_der_Vaart2015-xk}, but these procedures can lead to overfitting the model to the calibration data due to neglecting the conditional and stochastic aspects of data-generation and observation \citep{Brown2005-fq}. 

ABM calibration can be complicated by path-dependence and nonlinearities resulting from feedbacks, which increase the conditional nature of observations. To avoid overfitting and account for the stochastic elements of a model, another approach is to choose a model structure and parameter values which are most probable given the observations and prior information about system dynamics \citep{Jaynes2003-bw, Robert2007-rf}. Another complication of ABM development is the risk of over-parameterization. Over-parameterized models, which include variables and dynamics outside of what is supported by the evidence, may be an additional reason for overfitting observations and generalize less well for the purposes of prediction while adding minimal theoretical benefit. However, whether descriptive ABMs are intended to be used for explanation or prediction, these features suggest a need for quantification of model and parametric uncertainty, as observed patterns may be contingent on stochastic forcings or particular initial conditions. In particular, probabilistic calibration takes into account the stochastic nature of the data by modeling the likelihood of the observations given the external forcings or internal interactions and feedbacks.

This discussion leads to a simple overarching question: How much data is required to probabilistically calibrate agent-based models? We address this question using a Bayesian approach to uncertainty quantification, based on the Bayesian interpretation of parameter values as random variables. We focus on three questions. First, how do varying dataset sizes affect the statistical calibration of an ABM? Increasing complexity of agent decision rules (in the sense of the number of parameters) and including agent-agent and agent-environment interactions and feedbacks (in the sense of emergence) can reduce the ability to constrain model parameters or test hypotheses, particularly when data may be relatively scarce. Second, can we distinguish between models with varying levels of complexity, either in terms of high-dimensional decision rules or the types of agent interactions with each other and the environment? Third, how are calibration and hypothesis-testing affected by the use of spatially-aggregated data (as opposed to observations of individual agents), which may be all that are available due to data-collecting limitations or considerations of anonymity?

For a concrete example, we focus on the particular problem of modeling housing abandonment under flood risks, following the excellent work of  \citet{Tonn2017-ro}. Housing abandonment poses potentially severe economic problems for settlements along rivers and coastlines \citep{Fowler2017-yh}. Residents who haven’t experienced flooding themselves may abandon their homes if their neighbors do due to depreciating values or anticipation of future flooding. An associated ABM, and two nested submodels with fewer interactions and feedbacks, are illustrated by the influence diagrams in Figure~\ref{fig:influence}. 

\section{Methods}

\subsection{Models}
\label{sec:models}
\begin{figure}
\centering
\includegraphics[width=.95\textwidth]{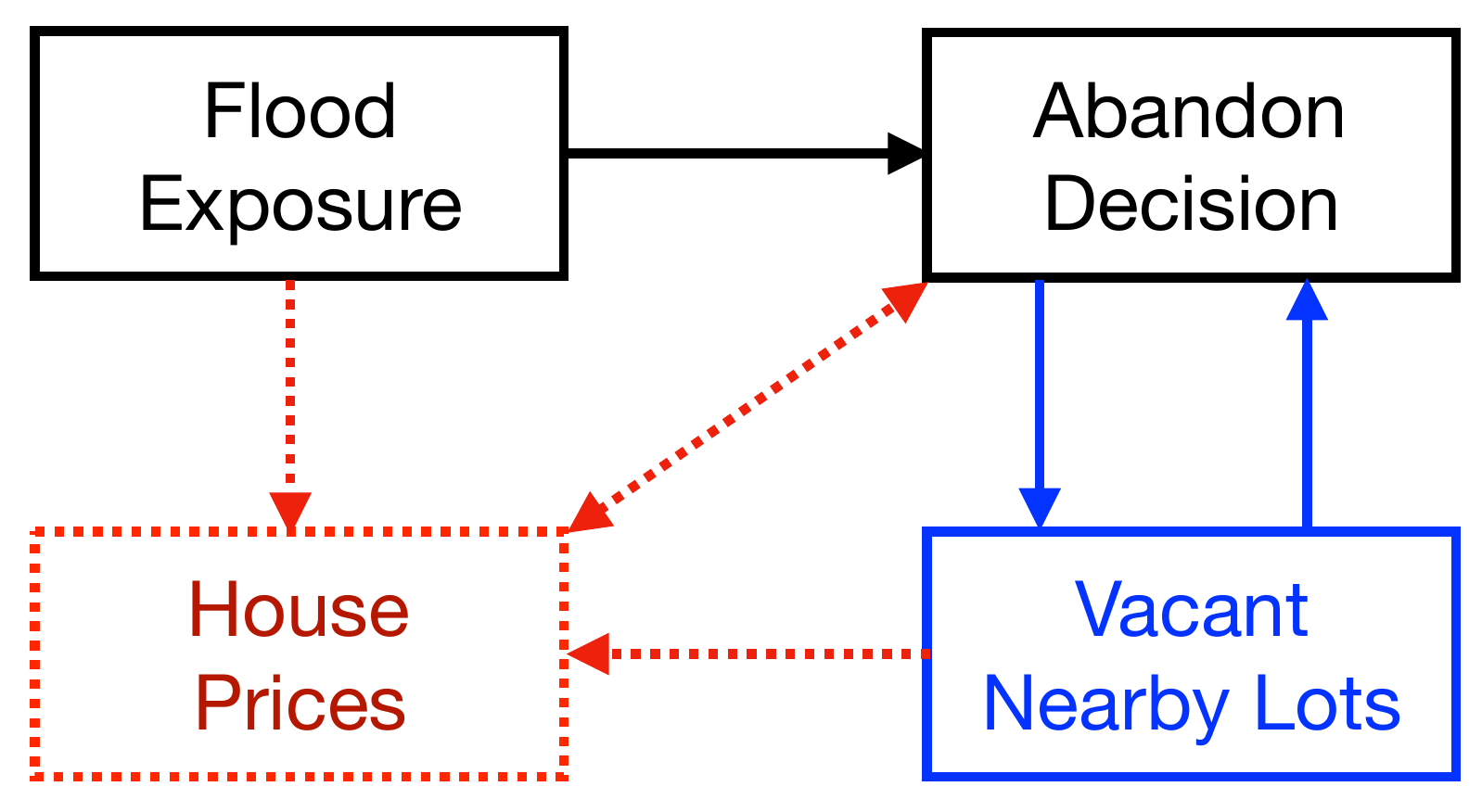}
\caption{Influence diagrams  of three nested ABMs for housing abandonment decisions under evolving flood pressure. The black components form a basic model without interactions (the ``no interaction'' model), in which abandonment decisions are based only on floods experienced by each agent. The blue and black components form a model with spatial interactions (the ``spatial interactions'' model), in which agents move due to experienced floods and the proportion of neighboring lots which are vacant. For context, we also show a more complex model (with the red additions) with spatial and economic interactions (the ``economic interactions'' model), in which housing market dynamics are affected by abandonments and floods and the abandonment decision includes housing values.}
\label{fig:influence}
\end{figure}

We analyze two ABMs (represented by the influence diagram in Figure~\ref{fig:influence}). In both models, agents decide to vacate their homes using a probabilistic decision process (logistic regression), as opposed to maximizing utility or using heuristics (which are more common in ABMs in certain application areas, such as land use \citep{Groeneveld2017-tx}. Once a house is abandoned, there is a chance that it is occupied by a new agent in a subsequent year. 

In the first ("simple") model, the probability of housing abandonment is determined only by the frequency of experienced floods over the previous ten years, is the ``no-interactions'' model, and is determined by three parameters: the logistic regression intercept $\beta_0$, the logistic regression coefficient for flood frequency $\beta_1$, and the probability that vacant houses are filled by a new agent $\alpha$.  Denote by $s^i_t$ the state of parcel $i$, where $s^i_t=1$ if the parcel is occupied and $0$ if the parcel is vacant. The probability that this parcel switches from occupied to abandoned is 
\begin{equation*}
    p\left(s^i_t = 0 | s^i_{t-1} = 1\right) = \text{logit}^{-1}\left(\beta_0 + \beta_1 r_t\right),
\end{equation*}
where $r_t$ is the frequency of flooding events for the previous ten years. The probability that this parcel switches from abandoned to occupied is constant, 
\begin{equation*}
    p\left(s^i_t = 1 | s^i_{t-1} = 0\right) = \alpha.
\end{equation*}

In the second (``spatial-interactions'') model, we include an additional logistic regression covariate, the fraction of neighboring plots which are vacant at the start of time $t$. As a result, this model has four parameters, including the coefficient for this neighboring-vacancy covariate. These state transition equations give both models a Markovian structure.

Of course, these models could be further expanded, for example by including housing market dynamics (see Figure~\ref{fig:influence}). In this study, we focus for clarity just on the two models, as we are interested in understanding the process and results of statistical calibration for simple ABMs. 

\subsection{Data}
\begin{figure}
    \centering
    \includegraphics[width=.95\textwidth]{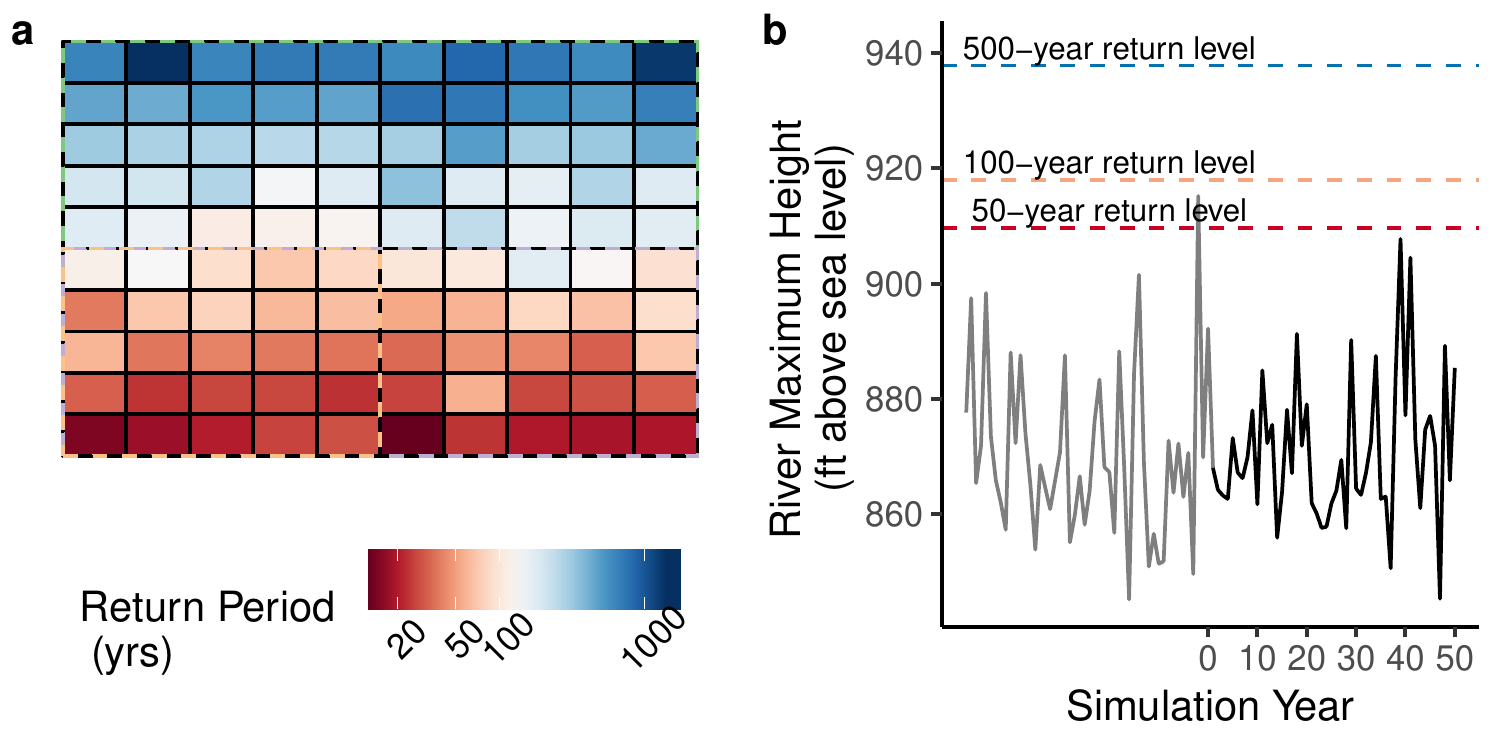}
    \caption{Flood information for the synthetic housing domain and river used in this study. Subfigure a) are the flood return periods for each parcel. The dashed outlines correspond to the sub-domains used in experiments with differing numbers of agents: orange is the 25-agent domain, purple is the 50-agent domain, and green is the 100-agent domain. Subfigure b) show the maximum annual river heights, for the 50-year period prior to the start of the simulations and for the 50 years used in the maximum-length simulation.}
    \label{fig:setting}
\end{figure}

\begin{figure}
    \centering
    \includegraphics[width=.95\textwidth]{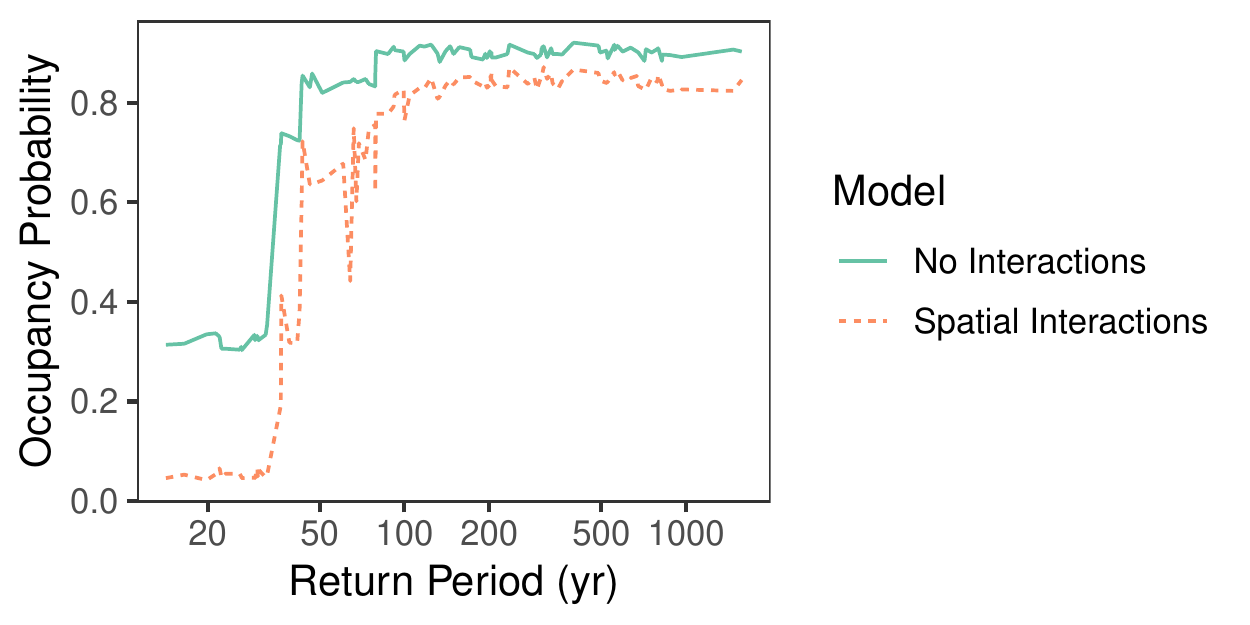}
    \caption{Occupancy probability (over 1000 sample realizations) for each parcel after a 50-year model run for the no-interactions model and the spatial-interactions model. The simulated model used observations of 100 parcels.}
    \label{fig:compare}
\end{figure}

We use the models described in Section~\ref{sec:models} in a perfect model experiment (see, for example, \citet{Olson2013-pe} or \citet{Reed2012-vw}, so that the data-generating process and model parameters are known. We generated pseudo-observations for the perfect model experiment using the model with spatial interactions, to see if we could successfully test for this effect. The pseudo-observations are generated using the spatial-interactions model for an artificial riparian settlement and realizations of annual flood height maxima from a generalized extreme value distribution. The parcel return periods and river heights are shown in Figure~\ref{fig:setting}. The additional dynamic mechanism resulting from spatial interactions leads to increased probabilities of parcel abandonment for all return periods across realizations of the stochastic process, even for parcels that are far from floods (Figure~\ref{fig:compare}).

Parcel residency was initialized by assuming that each parcel had a 99\% probability of having a resident in year 0. We used varying combinations of observed years and parcels (see Figure~\ref{fig:setting}). The combinations were 10, 25, and 50 years, and 25, 50, and 100 parcels. Annual maxima river heights were simulated from a generalized extreme value distribution with location parameter 865, scale parameter 11, and shape parameter 0.02. Data-generating parameter values were -6 for the logistic intercept, 20 for the local-flood coefficient, 4 for the neighboring-vacancy coefficient, and 0.01 for the vacancy-fill probability.

As data may not be available in individualized forms, we examine the power of data for calibration and hypothesis testing about model structures in both individual and aggregate forms. In the individual case, the data set contains observations of each observed parcel at each time. In the aggregate case, we observe the total number of abandoned parcels at each time.

\subsection{Calibration}

We use a Bayesian framework for model calibration, based on Bayes' Theorem \citep{Bayes1763-at}:
\begin{equation*}
p(\theta | y) \propto p(y | \theta) p(\theta),    
\end{equation*}
where $p(\theta | y)$ is the posterior density, $p(y | \theta)$ is the data likelihood, and $p(\theta)$ is the prior.

We constructed the priors that were used in this study (Table~\ref{tab:priors}) using a rough understanding of the model dynamics, so that the resulting probabilities of abandonment seemed plausible. They are intentionally not centered on the known data-generating parameter values. 

\begin{table}
    \centering
    \begin{tabular}{l|r}
    Parameter & Prior Distribution \\
    \hline
    Intercept & Normal(-7, 1)\\
    Flood Coefficient & Normal(19, 2)\\
    Vacancy Coefficient & Normal(5, 2)\\
    Vacancy Fill Probability & Beta(1, 10)
    \end{tabular}
    \caption{Prior distributions for each of the parameters in the no-interactions and spatial-interactions models.}
    \label{tab:priors}
\end{table}

For both individual-parcel and aggregate data, we model the probability of each parcel being vacant and compute the appropriate likelihood, treating each parcel’s vacant status at time $t$ as independent and identically distributed conditional on the state in time $t-1$. This representation (marginalizing over agent states to represent the model dynamics as a Markov chain) is common for many ABMs \citep{Izquierdo2009-sf}. In the individual data case, we use a binomial likelihood for each parcel at each time, with the probability of a vacant parcel determined using the Markovian representation after marginalizing. 
In the aggregate data case, we use a Poisson likelihood, which is commonly used to model count numbers, on the expected number of vacant parcels. For more details on how these likelihood functiosn are specified, see Supplemental Section S1.

The models described above are fast enough to use Markov Chain Monte Carlo (MCMC) for the Bayesian inversion. MCMC is an extremely general method for sampling from the posterior distribution and has previously been used for calibrating ABMs \citep{Keith2013-xg}. We use 150,000 Metropolis-Hastings \citep{Hastings1970-zq} iterations after a preliminary adaptive run \citep{Vihola2012-ml} of 30,000 iterations, which is used to estimate the starting value of the production run as well as the MCMC sampling distribution. The preliminary run is initialized at the maximum-likelihood estimate.

While our choosen method of Metropolis-Hastings MCMC has the ability to produce high-fidelity approximations to the full joint probability distribution of the model parameters \citep{Robert2014-yf}, it may be computationally intractable for complex models featuring long runtimes or high-dimensional parameter spaces. An additional complication is the need to specify a statistical likelihood function, which may be difficult for particular applications. In general, there is a trade-off between computational speed and accuracy of the resulting parameter distributions. Some alternative approaches to statistical calibration of ABMs, which are aimed at reducing computational requirements or likelihood specification, include statistical emulation  \citep{Lamperti2018-sl, Oyebamiji2017-ex}, particle filtering \citep{Kattwinkel2017-ct}, and approximate Bayesian computation \citep{Fabretti2018-gd, Siren2018-tf, Van_der_Vaart2015-xk, Van_der_Vaart2016-er}. While these methods reduce the computational burden, they come at a cost of potentially severe statistical approximations that can influence the parameter estimates \citep{Frazier2018-pw, Kunsch2013-et, Robert2011-pf, Singh2018-qq}.

\subsection{Model Selection}
We estimate the marginal likelihoods for each model  using the bridge sampling \citep{Meng1996-en}. The importance density is a truncated multivariate normal with mean and covariance derived from the MCMC output. The truncation occurs along the vacancy fill probability dimension, to ensure that this parameter only takes values between zero  and one. We use 5,000 posterior and importance samples in the bridge sampling estimator, which results in standard errors \citep{Fruhwirth-Schnatter2004-td} for the log-marginal likelihoods of orders of magnitude smaller than 1e-3. 

\section{Results}

\subsection{Calibration}

\begin{figure}
    \centering
    \includegraphics[width=1\textwidth]{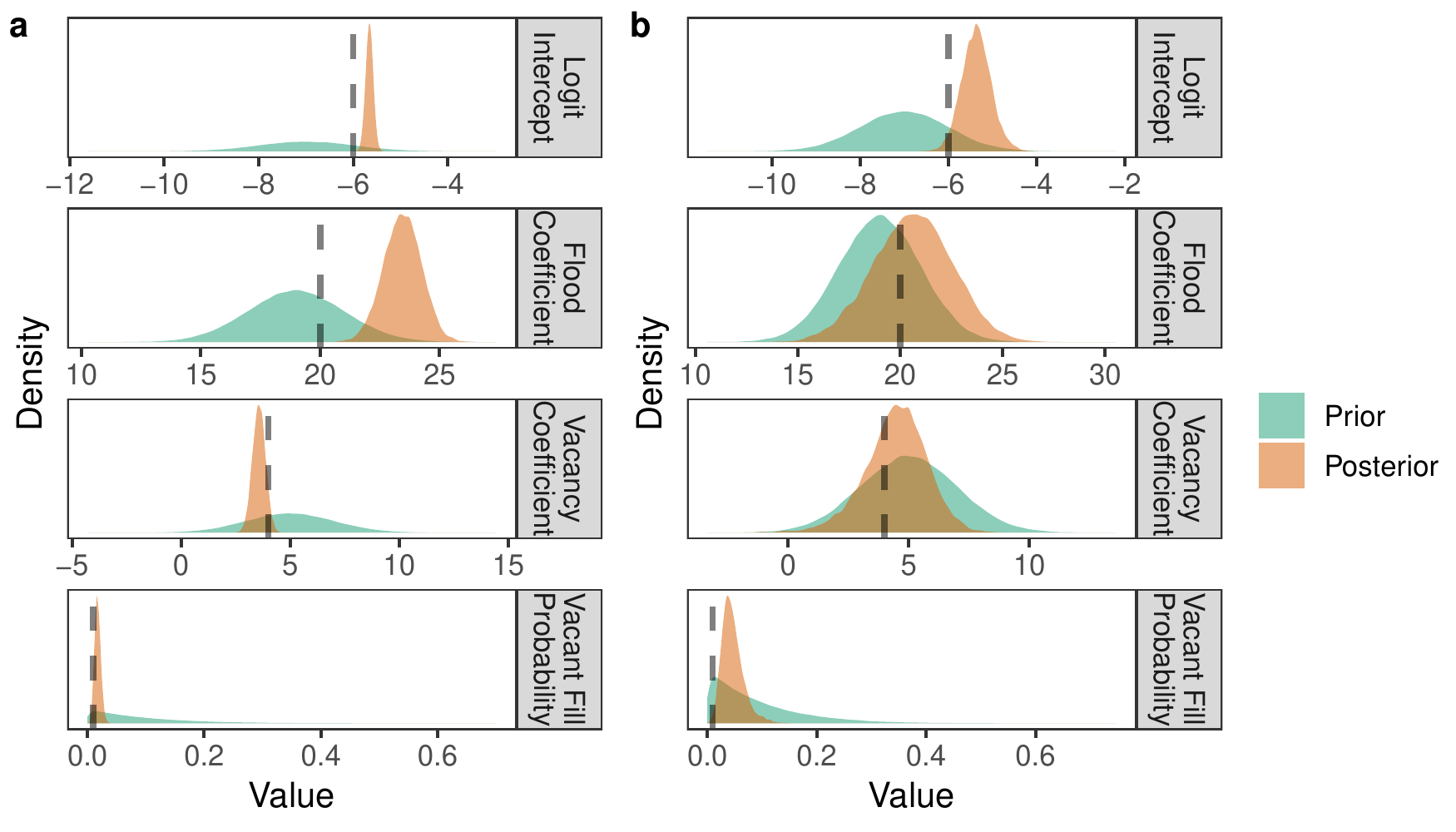}
    \caption{Calibration results (prior and posterior distributions) for the spatial-interactions model after assimilating 50 years of observed data with 100 observed parcels. The dashed vertical line is the value used in the data-generating process. Panel a) is after assimilating individual-parcel data, and panel b) is after assimilating aggregated data.}
    \label{fig:post}
\end{figure}

The structure of the data (individual-parcel versus spatially-aggregated) strongly influences the final shape of the posterior distribution, both due to the number of data points and the different likelihood function specifications. Figure~\ref{fig:post} shows the result of updating the prior distributions (specified in Table~\ref{tab:priors}) with 50 years of pseudo-observations of 100 parcels. For certain key parameters (such as the logistic regression coefficient for the local flooding frequency), aggregated data (the total number of abandoned parcels at each time) leaves the posterior close to the prior (Figure~\ref{fig:post}b). For individual parcel data,  while the marginal posterior is sharpened much further (Figure~\ref{fig:post}a). 

\begin{figure}
    \centering
    \includegraphics[width=1\textwidth]{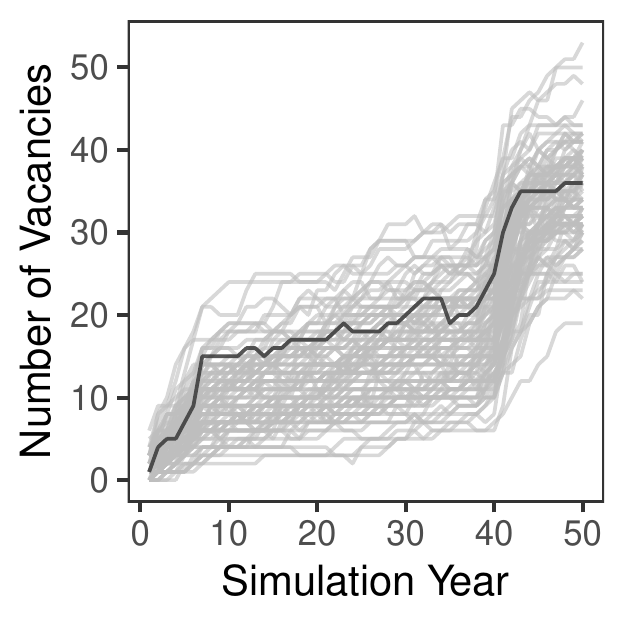}
    \caption{Number of vacant parcels for one hundred sample realizations from the complex model, with spatial interactions. The simulated model used observations of 100 parcels for 50 years. The black line is the realization used for that calibration experiment in this study.}
    \label{fig:realizations}
\end{figure}

While it appears from Figure~\ref{fig:post}a that the original decision rules are not fully recovered (looking at the posterior density at the data-generating value), it is important to keep in mind the influence of stochasticity in the realized data. Running the same model with the same parameters can yield model output with very different dynamics due to stochastic forcings, particularly in the presence of high levels of path dependence and positive feedbacks (see Figure~\ref{fig:realizations}). Between the strong influence of the stochastic elements in the model and the relative lack of sensitivity of the logistic regression to parameter values close to the data-generating value, it is not necessarily surprising that the data-generating value is assigned a relatively low density. 

\begin{figure}
    \centering
    \includegraphics[width=.95\textwidth]{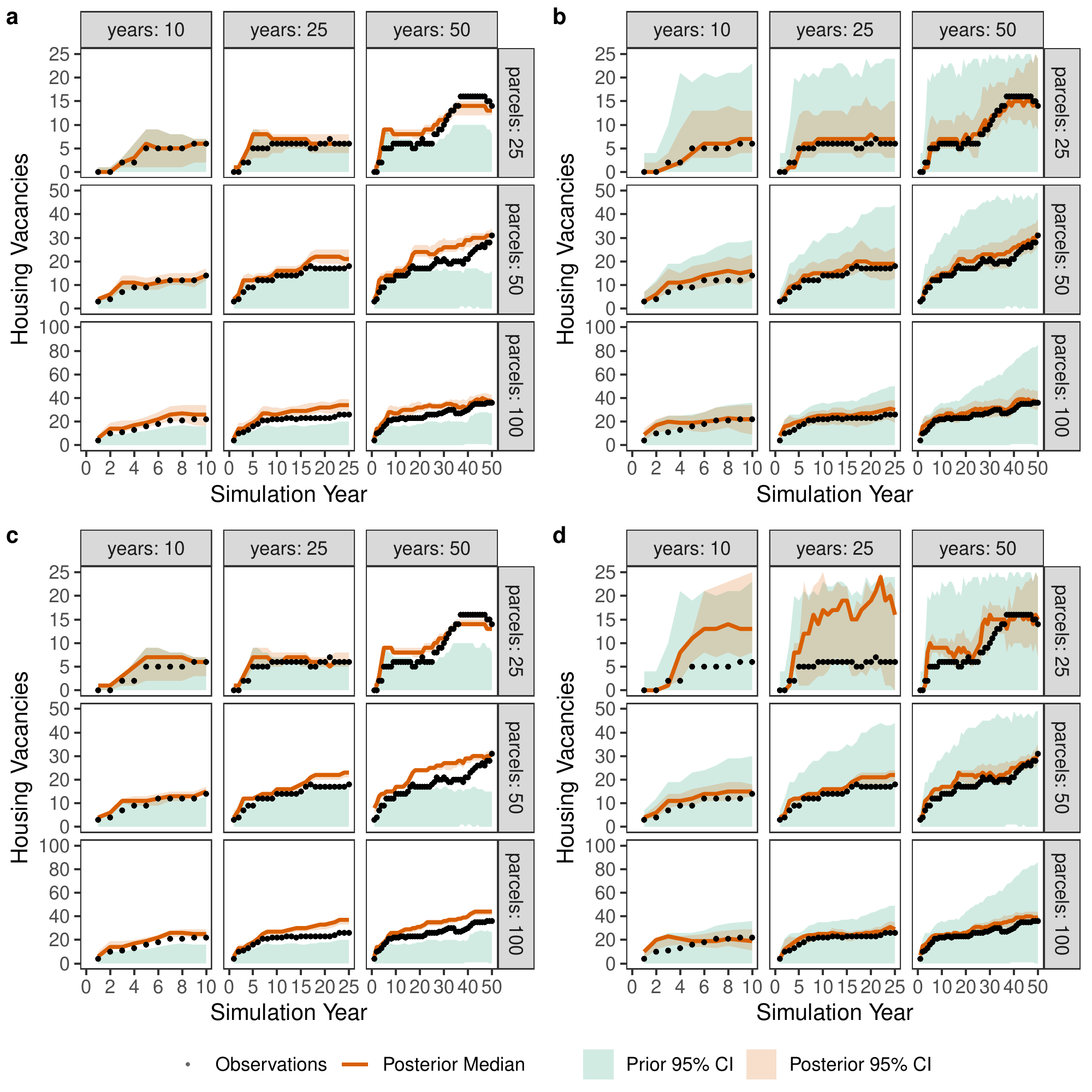}
    \caption{Prior and posterior hindcasts of the number of vacant parcels for the no-interactions model with aggregated data (panel a), the spatial-interactions model with aggregated data (panel b), the no-interactions model with individual-parcel data (panel c), and the spatial-interactions model with individual-parcel data (panel d) for varying combinations of observed years and parcels.}
    \label{fig:hindcast}
\end{figure}

The full posterior parameter estimates feature a high degree of correlation between parameters. For example, the two logistic regression coefficients for flood frequency and proportion of neighboring abandoned parcels have a correlation coefficient of -0.66: a lower sensitivity to experienced floods can be offset by an increased sensitivity to neighbor behavior. Another example is the high positive correlation between the probability of a vacant lot being re-occupied and both the logistic regression intercept term and the coefficient for neighboring parcels (r=0.73 in both cases). Similar interactions would be missed by a deterministic calibration combined with one-at-a-time sensitivity analysis \citep{Ten_Broeke2016-bc}. 

To validate the calibrated model, we analyze the hindcasting ability of the posterior predictive distribution (shown in Figure~\ref{fig:hindcast}). While the three-parameter no-interactions model is well constrained by smaller data sets, the poor fit of the posterior predictive distribution compared to the pseudo-observations for increased amounts of data reveals the missing abandonment dynamic mechanism. Without spatial interactions, the no-interactions model calibration results in a higher sensitivity to experienced flooding to account for the data, which results in an overestimate of the number of abandoned parcels in later years. Meanwhile, the spatial-interactions model, which has one additional parameter, requires more data to constrain the model (25 observed parcels is insufficient with up to 50 years of data), but, once constrained, fits the pseudo-observations better than the no-interactions model. In general, having a larger spatial domain/numbers of agents facilitates calibration more than having a longer data record. 

While Figure~\ref{fig:hindcast} might appear to show that calibration with aggregate data results in a better fit to the observations, this is likely an artifact of two different components of the modeling process. First, the likelihood functions used for the aggregate data calibration was different than in the individual-data case (Poisson vs. binomial), which will change the shape of the posterior distribution. Second, in the aggregate case, we calibrated the model directly against the aggregated counts  shown in Figure~\ref{fig:hindcast}, while in the individual-data case, the expected state of each parcel was used. These two differences make it difficult to compare the quality of the hindcast, as they are structurally different calibration procedures. We would expect that in the latter case, there is greater uncertainty about the total count of abandoned parcels. We do not view this is not a flaw with the individual-data procedure, and it likely better reflects the true underlying uncertainty, given the complex dynamics in the model.

\subsection{Model Selection}
More complex ABMs can be thought of as being constructed by adding new interactions and feedbacks to simpler ABMs, as illustrated in Figure~\ref{fig:influence}. This allows us to view this type of model selection as hypothesis testing for the presence of additional feedback mechanisms \citep{Cottineau2015-fd}. One standard method of comparing the fit of Bayesian models to data is by computing Bayes factors \citep{Kass1995-yr}. The Bayes factor is the ratio of marginal likelihoods of two models (the integral of the data likelihood over the posterior). Posterior model structural probabilities can be calculated by combining prior beliefs about the relative probability of the competing models with the Bayes factor.

One important consideration when using Bayes factors is the role of the prior in the computation \citep{Robert2007-rf}, particularly when they are used for point-null hypothesis testing. Here, we use the same priors for corresponding parameters  to reduce  this effect.

\begin{figure}
    \centering
    \includegraphics[width=.7\textwidth]{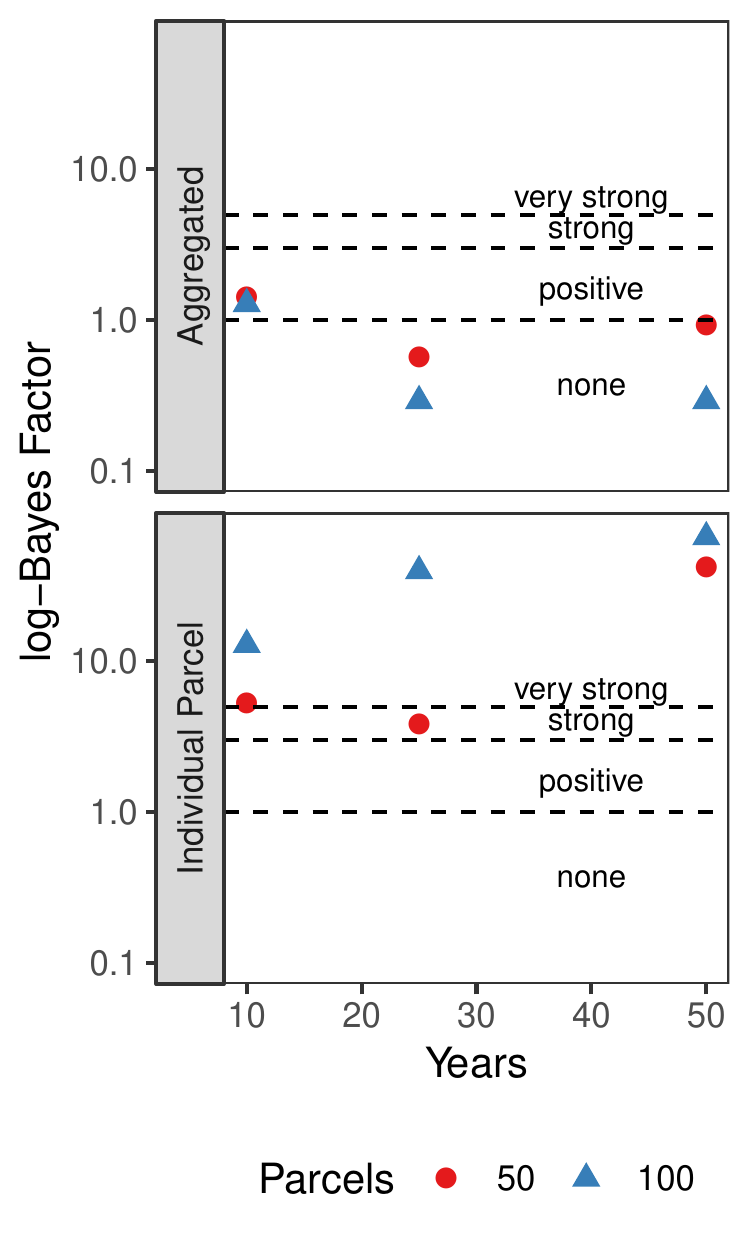}
    \caption{Log-Bayes Factors when comparing the with-interactions model to the no-interactions model for 50 and 100 parcels and 10, 25, and 50 years. The thresholds for varying degrees of evidence are taken from \citet{Kass1995-yr}. The marginal likelihoods for each model were estimated using bridge sampling \citep{Meng1996-en} with 5000 samples and a truncated multivariate normal importance density.}
    \label{fig:bf}
\end{figure}

For our perfect model experiment, we would expect additional (in terms of the number of observations) and spatially explicit (rather than aggregated) data to improve the ability to distinguish between the data-generating spatial-interactions model and the simpler no-interactions model. In Figure~\ref{fig:bf}, we show the log-Bayes factors (along with thresholds for evidence levels proposed by \citet{Kass1995-yr}) to summarize the evidence for the spatial-interactions model versus the no-interactions model. We neglect the case with 25 observed parcels due to unreasonably high estimates, likely due to the ill-constrained spatial-interactions model (however, the hindcasts in Figure~\ref{fig:hindcast} shows the qualitatively better fit of the no-interactions model for this data set, particularly in the individual-data scenario). 
For individual-parcel data, with more than 25 observed parcels, there is at least strong evidence for the spatial-interactions model no matter how long the parcels were observed, which confirms the qualitative assessment (on the summary statistic of total abandoned parcels) obtained by comparing the hindcasts in Figures~\ref{fig:hindcast}c and~\ref{fig:hindcast}d.

On the other hand, when aggregated data is used for calibration, there is essentially no quantitative evidence for the spatial-interactions model. This is the case whether we compare the models using Bayes factors or a predictive information criterion such as the Watanabe-Akaike information criterion (\citep{Watanabe2010-kb, Vehtari2017-ln}, estimated (along with standard errors of the differences) using 10,000 posterior samples. Predictive model comparison methods avoid the direct influence of the prior on the comparison and allows for an intuitive comparison between models which have different parameterizations \citep{Gelfand1998-gz}. The one-standard error range of the difference in WAIC between the spatial-interactions and the no-interactions model is between -2 and 2, which can be interpreted as no difference in support between the two models \citep{Burnham2004-do}. However, a qualitative assessment obtained by comparing Figures~\ref{fig:hindcast}a and~\ref{fig:hindcast}b might lead a modeler to conclude that the spatial-interactions model fits the observations better than the no-interactions model. This suggests that hindcasting  can serve an important supporting role to quantitative model selection.

\section{Discussion}

Probabilistic calibration is an important component of the descriptive agent-based modeling process due to the influence of stochastic noise via path-dependence and feedback loops (as illustrated in Figure~\ref{fig:realizations}). However, as our results illustrate, each additional parameter can considerably increase the calibration data requirements. Trying to include every hypothesized feedback mechanism in the final model choice, without supporting evidence, can pose problems from statistical as well as a decision-theoretical points of view \citep{Box1976-vy, Jaynes2003-bw, Robert2007-rf}. Starting with a simple model and adding complexity when supported by the data can produce more skillful hindcasts, projections, and more powerful insights \citep{Box1979-bk, Holling1966-rl}. 

An additional concern is the specification of prior distributions. When less data is available (particularly in summarized or aggregated form), that data will have less power to update the prior distributions.This suggests that priors should be as informative as possible (with a strong warning that priors ought not to be more informative that can be supported). While we did not take prior correlations between parameters into account for this experiment, good priors for real-world problems will include prior information about correlations between parameters.

One approach to creating informed priors which include information about the relationships between parameters is probabilistic inversion \citep{Kraan2000-ce, Fuller2017-ie}, in which expert assessments (or, as an alternative, the results of judgement and decision-making or economic experiments) can be used to update more generic priors in a way which is consistent with those assessments or experimental results. This allows the survey or experimental participants to provide information directly about outcomes rather than about model parameters, and allows for a separation of the data involved in the prior construction and Bayesian updating processes.

\section*{Acknowledgements}

The authors would like to thank Ben S. Lee, Joel Roop-Eckart, Tony E. Wong, and Skip Wishbone for their input and contributions. This work was partially supported by the National Science Foundation (NSF) through the Network for Sustainable Climate Risk Management (SCRiM) under NSF cooperative agreement GEO-1240507, the Penn State Center for Climate Risk Management, and by the U.S. Department of Energy, Office of Science, Biological and Environmental Research Program, Earth and Environmental Systems Modeling, MultiSector Dynamics, Contract No. DE-SC0016162. Any opinions, findings, and conclusions or recommendations expressed in this material are those of the authors and do not necessarily reflect the views of the NSF. All codes for pseudo-data generation, model analysis and figure generation can be found at \href{http://www.github.com/vsrikrish/ABM/tree/calibration}{http://www.github.com/vsrikrish/ABM/tree/calibration}.

\section{Author contributions statement}

V.S. and K.K. conceptualized the research. V.S. wrote the model and analysis codes. V.S. and K.K. designed the figures and wrote the paper.

\section*{Additional information}

\textbf{Competing interests}: The authors are not aware of competing interests. 

\bibliography{references}

\end{document}


\maketitle

\section{Likelihood Functions for Models}

The likelihood functions used in this study are derived using the Markov property. A stochastic process has the Markov property if the present state $X_t = x_t$ contains all relevant information for computing the conditional probability of the future states on the past and present states. In other words, a process has the Markov property if
\begin{equation*}
    p(X_{t+1} = x_{t+1} | X_1 = x_1, \ldots, X_t = x_t) = p(X_{t+1} = x_{t+1} | X_t = x_t).
\end{equation*}
The models described in this paper have the Markov property because the occupied or unoccupied status of a parcel $i$ at time $t+1 $ depends only on the current (time $t$) state of the parcel, where the augmented state $X^i_t$ is described by the following variables:
\begin{itemize}
    \item $s^i_t$: house occupancy indicator ($s^i_t = 0$) or unoccupied ($s^i_t = 1$);
    \item $r^i_t$: historical flood rate (based on the previous ten years or the residency time of the current owner, whichever is larger);
    \item $v^i_t$: fraction of vacant neighboring parcels (for the spatial interactions model only).
\end{itemize}
The flood rate $r^i_t$ is updated based on the exogenous extreme water level and the elevation of parcel $i$ as well as the occupancy indicator (it is reset to 0 when the parcel switches from occupied to unoccupied, and remains at zero until the parcel is re-occupied). The vacant neighboring parcel fraction $v^i_t$ is updated based on the occupancy indicator $s_{t-1}$ for the neighboring parcels. The occupancy indicator is then updated probabilistically, with the conditional probability of an occupied lot becoming vacant given by
\begin{equation}
    P^{10}_t = p(s^i_t = 0 | s^i_{t-1} = 1) = \text{logit}^{-1}(y^i_t),
\label{eq:prob1}
\end{equation}
where $y^i_t = \beta_0 + \beta_1 r^i_t$ for the no-interactions model and $y^i_t = \beta_0 + \beta_1 r^i_t + \beta_2 v^i_t$ for the spatial interactions model, and $\beta_i$ are the regression intercept and coefficients. The conditional probability of an unoccupied parcel becoming occupied is
\begin{equation}
    P_t^{01} = p(s^i_t = 1 | s^i_{t-1} = 0) = \alpha.
\label{eq:prob2}
\end{equation}

Combining the conditional probabilities in Equations \eqref{eq:prob1} and \eqref{eq:prob2} yields a recursive updating equation,
\begin{equation*}
\begin{pmatrix}p(s^i_t = 0) \\ p(s^i_t = 1)\end{pmatrix} = \begin{pmatrix}1-P{01} & P_{01} \\ P_{10} & 1 - P_{10}\end{pmatrix} \begin{pmatrix}p(s^i_{t-1} = 0) \\ p(s^i_{t-1} = 1)\end{pmatrix}.
\end{equation*}
Denoting 
\begin{equation*}
    P_t = \begin{pmatrix}1-P^{01}_t & P_{01} \\ P_t^{10} & 1 - P_t^{10}\end{pmatrix},
\end{equation*}
we can write the unconditional probabilities for $s_i^t$ (unconditional as we view the initial probability of vacancy as fixed, $p(s_0^i = 1) = 0.99$) as
\begin{equation}
    \begin{pmatrix}p(s^i_t = 0) \\ p(s^i_t = 1)\end{pmatrix} = \prod_{\tau=1}^t P_\tau \begin{pmatrix}p(s^i_0 = 0) \\ p(s^i_0 = 1)\end{pmatrix}.
    \label{eq:uncondprob}
\end{equation}

The probability distribution which defines the likelihood function depends on whether we assume we have individual-parcel or aggregated data. In the individual-parcel case, we know the occupancy state of each parcel at each time. We use a binomial distribution (which models the probability of a Boolean outcome) at each time using the unconditional probabilities defined by Equation~\eqref{eq:uncondprob} for each parcel, and assume that the likelihoods for each parcel at each time are independent, as the dependence on neighboring parcels (which only exists in the spatial-interactions model) is captured in Equation~\eqref{eq:uncondprob} via $v_i^t$. 

For the aggregated case, we only know the total number of vacant parcels at each time, and we compute the expected number of vacant parcels by summing the vacancy probability over each parcel, $\lambda_t = \sum_i p(s^i_t)$. We then model this count using a Poisson distribution with parameter $\lambda_t$.